\documentclass[twocolumn,showpacs,
amsmath,amssymb]{revtex4}
\usepackage{amssymb}
\usepackage[dvips]{graphicx}
\begin{document}

\title{Theory of quantum magneto-oscillations in underdoped cuprate superconductors}
\author{ A. S. Alexandrov}

\affiliation{Department of Physics, Loughborough University,
Loughborough LE11 3TU, United Kingdom\\}

\begin{abstract}
Solving the Gross-Pitaevskii-type equation it is shown that the
magneto-oscillations,
 observed in the superconducting state of a few
underdoped
 cuprates,
 originate in the quantum interference of the vortex lattice with
 nanoscale crystal lattice modulations of the order parameter as revealed
 by Scanning Tunnelling Microscopy (STM). The commensuration of the vortex lattice and crystal lattice have $1/B^{1/2}$ periodicity, rather than  $1/B$ periodicity of conventional normal state
 magneto-oscillations.  Experimental conditions allowing for a resolution
 of two different types of quantum magneto-oscillations are
 outlined.
\end{abstract}


\maketitle

Until recently  no convincing signatures of
 quantum magneto-oscillations have been found  in
the normal state of cuprate superconductors despite significant
experimental efforts. There are no normal state oscillations even in
high quality single crystals of overdoped cuprates like
Tl$_2$Ba$_2$CuO$_6$, where conditions for
 de Haas-van Alphen (dHvA)  and Shubnikov-de Haas (SdH)
oscillations seem to be perfectly satisfied \cite{mac} and a large
Fermi surface is identified in the angle-resolved photoemission
spectra (ARPES)  \cite{plate}. The recent observations of
magneto-oscillations in  kinetic \cite{ley,ban} and magnetic
\cite{sin} response functions of underdoped YBa$_2$Cu$_3$O$_{6.5}$
and YBa$_2$Cu$_4$O$_8$ are perhaps even more striking since many
probes of underdoped cuprates including ARPES \cite{shen} clearly
point to a non Fermi-liquid normal state. Their description  in the
framework of the standard theory for a metal \cite{schoen} has led
to  a very small  Fermi-surface area of  a few percent of the first
Brillouin zone \cite{ley,sin,ban}, and to a
 low Fermi energy of only about the room temperature \cite{sin}.
Clearly such oscillations are incompatible with the first-principle
(LDA) band structures of cuprates, but might be compatible with a
non-adiabatic polaronic normal state of charge-transfer Mott
insulators \cite{aleadi}. Nevertheless their observation in  the
\emph{superconducting} (vortex) state well below the
$H_{c2}(T)$-line \cite{ley} raises a doubt concerning their normal
state origin.

Here I propose an alternative explanation of the
magneto-oscillations \cite{ley,ban,sin} as emerging from the quantum
interference of the vortex lattice and the checkerboard or lattice
modulations of the order parameter observed by STM with atomic
resolution \cite{stm}. The checkerboard effectively pins the vortex
lattice, when the period of the latter $\lambda= (\pi
\hbar/eB)^{1/2}$ is commensurate with the period of the checkerboard
 lattice, $a$. The condition $\lambda= Na$ , where $N$ is
a large integer, yields $1/B^{1/2}$ periodicity of the response
functions, rather than $1/B$ periodicity of conventional normal
state
 magneto-oscillations.

To illustrate the point one can apply the Gross-Pitaevskii (GP)-type
equation for the superconducting order parameter $\psi({\bf r})$,
generalized by us  \cite{alevor} for a charged Bose liquid (CBL),
since many observations including a small coherence length point to
a possibility that underdoped cuprate superconductors may not be
conventional Bardeen-Cooper-Schrieffer (BCS) superconductors, but
rather derive from the Bose-Einstein condensation (BEC) of
real-space pairs, such as  mobile small bipolarons \cite{ale96},
\begin{equation}
\left[E(-i\hbar {\bf \nabla}+2e {\bf A})-\mu+\int d{\bf r'} V({\bf
r} -{\bf r'})|\psi ({\bf r'})|^2 \right]\psi({\bf r}) =0.\label{gp}
\end{equation}
Here $E({\bf K})$ is the center-of-mass pair dispersion and the
Peierls substitution, ${\bf K} \Rightarrow -i\hbar {\bf \nabla}+2e
{\bf A}$ is applied with the vector potential ${\bf A}({\bf r})$.
\begin{figure}
\begin{center}
\includegraphics[angle=-90
,width=0.50\textwidth]{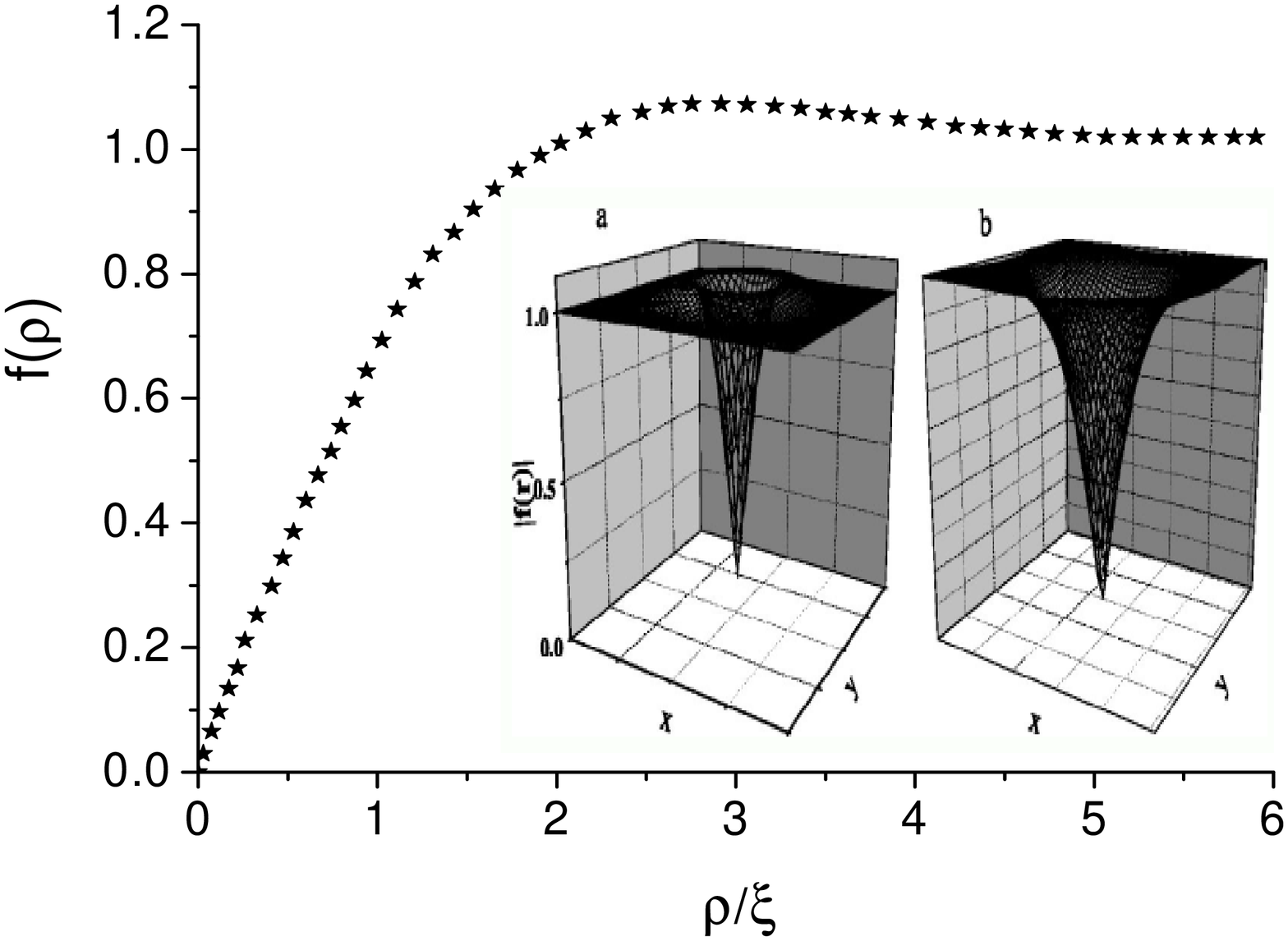} \vskip -0.5mm \caption{The order
parameter profile $f(\rho)=\psi({\bf r})/ n_s^{1/2}$ of a single
vortex in CBL \cite{alevor} (symbols). Inset: CBL vortex (a)
\cite{alevor,alekab} compared with the Abrikosov vortex (b)
\cite{abr} (here $\rho=[x^2+y^2]^{1/2}$). }
\end{center}
\end{figure}
The integro-differential equation (\ref{gp})
 is quite different from the  Ginzburg-Landau \cite{gin}
and  Gross-Pitaevskii \cite{gro} equations, describing the order
parameter in the BCS and neutral superfluids, respectively. In the
continuum (effective mass) approximation, $E({\bf K})=
\hbar^2K^2/2m^{**}$, with the long-range Coulomb repulsion between
double charged bosons, $V({\bf r})=V_c({\bf r})=4e^2/\epsilon_0 r$,
this equation describes a single vortex with a \emph{charged} core,
Fig.1, and the coherence
 length roughly the same as the screening radius, $
 \xi=(\hbar/2^{1/2}m^{**}\omega_{p})^{1/2}$.
  Here
$\omega_{p}=(16\pi n_s e^{2}/\epsilon_0 m^{**})^{1/2}$ is the CBL
plasma frequency,  $\epsilon_0$  the static dielectric constant of
the host lattice, $m^{**}$  the boson mass, and $n_s$ is the average
condensate density.  The chemical potential is zero, $\mu=0$, if one
takes into account the Coulomb interaction with a neutralizing
homogeneous charge background, or defines the zero-momentum
Fourier-component of $V_c({\bf r})$ as zero. Each vortex carries one
flux quantum, $\phi_0=\pi \hbar/e$ , but it has an unusual core,
Fig.1a, Ref. \cite{alevor}, due to a local charge redistribution
caused by the magnetic field, different from the conventional vortex
\cite{abr}, Fig.1b. Remarkably,  the coherence length turns out very
small, $\xi \approx 0.5$nm with the material parameters typical for
underdoped cuprates, $m^{**}=10 m_{e}$, $n_s=10^{21}cm^{-3}$ and
$\epsilon_{0} = 100$. The coherence length $\xi$ is so small at low
temperatures, that the distance between two vortices remains large
compared with the vortex size, $\lambda \gg \xi$, \cite{alekab} in
any  laboratory field reached so far \cite{ley,sin,ban}. It allows
us to write down the vortex-lattice order parameter, $\psi({\bf
r})=\psi_{vl}({\bf r})$, as
\begin{equation}
\psi_{vl}({\bf r})\approx n_s^{1/2} \left[1- \sum _{j} \phi({\bf r -
r}_j)\right],
\end{equation}
where $\phi({\bf r})=1-f(\rho)$, and  ${\bf r}_j=\lambda\{n_x,n_y\}$
with $n_{x,y}=0, \pm 1, \pm2, ...$ (if, for simplicity,  we take the
square vortex lattice). The function $\phi(\rho)$ is linear well
inside the core, $\phi(\rho)\approx 1- 1.52 \rho/\xi$ $(\rho\ll
\xi)$,  and it has a small negative tail, $\phi(\rho)\approx
-4\xi^4/\rho^4$  outside the core when $\rho \gg\xi$, Fig.1
\cite{alevor}.

In the continuum approximation with the Coulomb interaction alone
the magnetization of CBL follows the standard logarithmic law ,
$M(B) \propto \ln 1/B$ without any oscillations since  the magnetic
field profile is the same as in the conventional vortex lattice
\cite{abr}. However, more often than not the center-of-mass Bloch
band of preformed pairs, $E({\bf K}),$ has its minima at some finite
wave vectors ${\bf K}={\bf G}$ of their center-of-mass Brillouin
zone \cite{ale96,alebook}. Near the minima the GP equation
(\ref{gp}) is written as
\begin{eqnarray}
\left[{(-i\hbar {\bf \nabla}-\hbar{\bf G}+2e {\bf
A})^2\over{2m^{**}}}-\mu \right]\psi({\bf r})+ \cr \int d{\bf r'}
V({\bf r} -{\bf r'})|\psi ({\bf r'})|^2 \psi({\bf r}) =0
\label{gp2},
\end{eqnarray}
with the solution $\psi({\bf r})=\psi_{\bf G}({\bf r})\equiv e^ {i
{\bf G \cdot r}}\psi_{vl}({\bf r})$, if the interaction is the
long-range Coulomb one, $V({\bf r})=V_c({\bf r})$.
\begin{figure}
\begin{center}
\includegraphics[angle=-90
,width=0.48\textwidth]{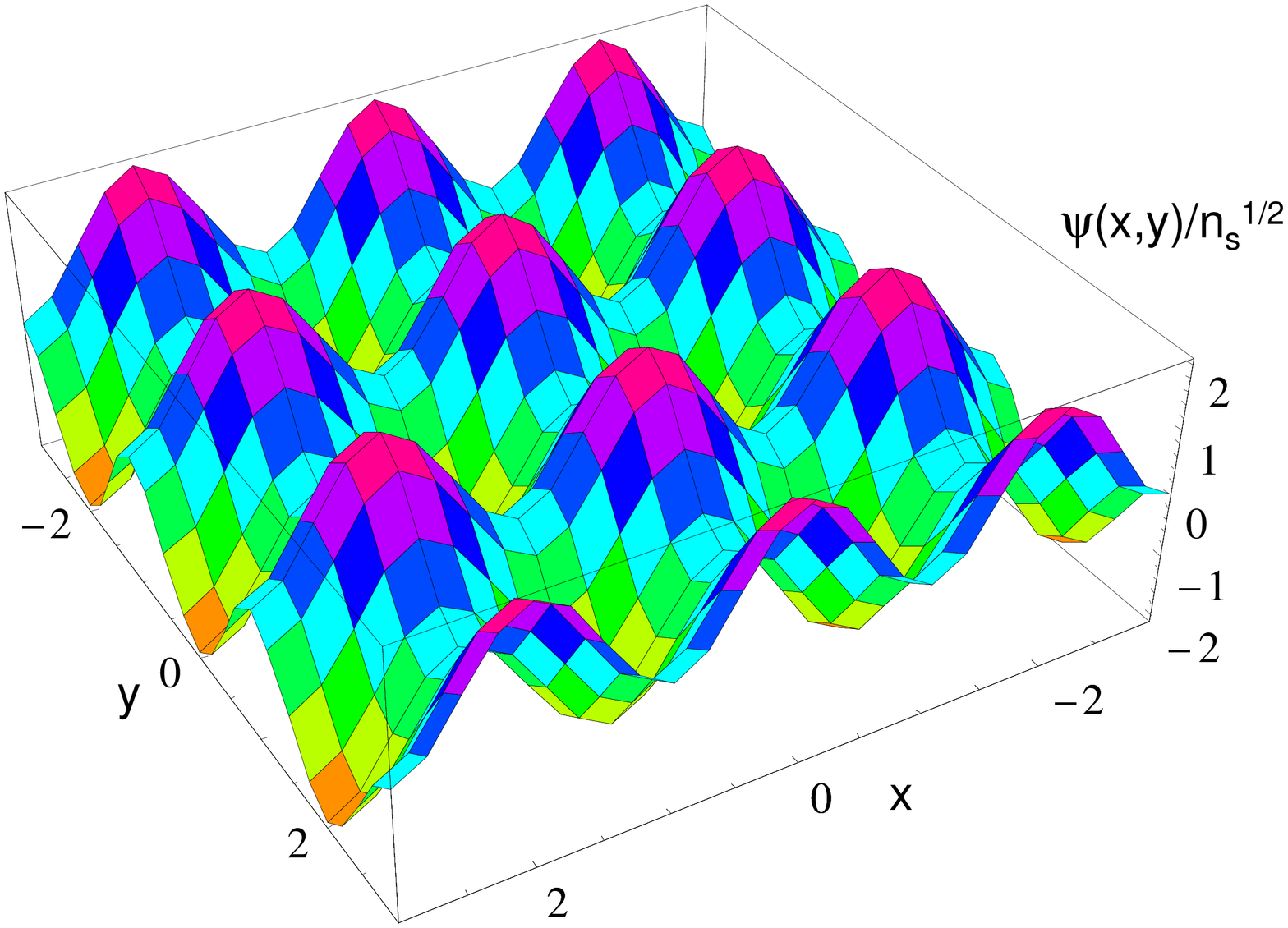} \vskip -0.5mm \caption{The
checkerboard d-wave order parameter of  CBL \cite{alesym} on the
square lattice in zero magnetic field (coordinates $x,y$ are
measured in units of $a$). }
\end{center}
\end{figure}
In particular, a nearest-neighbor (nn)  approximation for the
hopping of intersite bipolarons between oxygen p-orbitals on the
CuO$_2$ 2D lattice yields four generate states $\psi_{\bf G}$ with
${\bf G}_{i} =\{\pm 2\pi/a_0, \pm 2\pi/a_0\}$ , where $a_0$ is the
lattice period \cite{alebook}. Their positions in the Brillouin zone
move towards $\Gamma$ point beyond the nn approximation. The true
ground state is a superposition of four degenerate states,
respecting time-reversal and parity symmetries \cite{alesym},
\begin{equation}
\psi({\bf r})=An_{s}^{1/2}\left[ \cos (\pi x/a)\pm \cos (\pi
y/a)\right]\psi_{vl}({\bf r}). \label{two}
\end{equation}
Here we use the reference frame with $x$ and $y$ axes along the
nodal directions and  $a=2^{-3/2} a_0$. Two "plus/minus" coherent
states, Eq.(\ref{two}), are physically identical since they are
related via a translation transformation, $y \Rightarrow y+ a$.
Normalizing the order parameter by its average value $\langle
\psi({\bf r})^2\rangle=n_s$ and using $(\xi/\lambda)^2 \ll 1$ as a
small parameter  yield the following "minus" state amplitude,
$A\thickapprox 1-N\sum^{\infty}_{n=0}2
[\tilde{\phi}_1(2^{1/2}\pi/a)+\tilde {\phi}_2(2^{1/2}\pi/a)]
\delta_{n, R/2}+ [ \tilde {\phi}_1(2\pi/a)+\tilde {\phi}_2(2\pi/a)]
\delta_{n, R}$ for the square vortex lattice \cite{ref} with the
reciprocal vectors ${\bf g}=(2\pi/\lambda) \{n_x,n_y\}$. Here
$\delta_{n,R}$ is the Kroneker symbol, $R=\lambda/a$ is the ratio of
the vortex lattice period to the checkerboard period
($n=0,1,2,...$),  $N=BS/\phi_0$ is the number of flux quanta in the
area $S$ of the sample, and $\tilde{\phi}_k (q)= (2\pi/S)
\int_0^{\infty} d{\bf \rho} \rho J_0(\rho q) \phi^{k} (\rho) $ is
the Fourier transform of $k$'s power of $\phi(\rho)$, where $J_0(x)$
is the zero-order Bessel function.

The order parameter $\psi({\bf r})$, Eq.(\ref{two}) has the $d$-wave
symmetry changing sign in  real space, when the lattice is rotated
by $\pi /2$. This symmetry is
due to the pair center-of-mass  energy dispersion with the four minima at ${\bf %
K} \neq 0$, rather than due to  a  specific symmetry of the pairing
potential. It also reveals itself as a {\it checkerboard} modulation
of the carrier density with two-dimensional patterns  in zero
magnetic field,  Fig.2, as predicted by us \cite{alesym} prior to
their observations  \cite{stm}. Solving the Bogoliubov-de Gennes
equations with the  order parameter, Eq.(\ref{two}),  yields
 the real-space checkerboard
modulations of the single-particle density of states \cite{alesym},
similar to those observed by STM in cuprate superconductors.

Now we take into account that the interaction between composed pairs
includes a short-range  repulsion along with the long-range Coulomb
one, $V({\bf r})=V_c({\bf r})+v\delta({\bf r})$ \cite{alebook}. At
sufficiently low  carrier density the short-range repulsion is a
perturbation to the ground state, Eq.(\ref{two}), if the
corresponding characteristic length, $\xi_h= \hbar/(2m^{**}
n_sv)^{1/2}$ is  large compared with the coherence length $\xi$,
related to the long-range Coulomb repulsion, $\xi_h \gg \xi$. The
short-range repulsion constant $v$ is roughly the pair bandwidth $w$
of the order of $100$ meV times the unit cell volume, $v\approx w
a_0^3$ \cite{alebook}. Using this estimate one can readily show that
the perturbation treatment of the short-range interaction is
justified for any relevant density of pairs, if $\epsilon_0\lesssim
10^3$. On the other hand, a strong short-range interaction could
affect both the checkerboard and the vortex lattices, if $\xi_h$ is
comparable with $\xi$ and $a$.

Importantly the short-range repulsion energy of CBL, $U=(v/2)
\langle \psi({\bf r})^4 \rangle$, has a part, $\Delta U $,
oscillating with the magnetic field  as
\begin{equation}
{\Delta U\over{U_0}}\approx N  \sum_{n=0}^{\infty} \left[A_1
\delta_{n,R/2}+ A_2  \delta_{n,R}+A_3
\delta_{n,2R}\right],\label{part}
\end{equation}
where $U_0=v n_s^2/2$ is the hard-core energy of a homogeneous CBL,
and  the amplitudes are proportional to the Fourier transforms of
$\phi(\rho)$ as $A_1=15 \tilde{\phi}_1\left(2^{1/2}\pi/a\right)- 45
\tilde{\phi}_2\left(2^{1/2}\pi/a\right) +24
\tilde{\phi}_3\left(2^{1/2}\pi/a\right) -
6\tilde{\phi}_4\left(2^{1/2}\pi/a\right)+
8\tilde{\phi}_1\left(10^{1/2}\pi/a\right)-12\tilde{\phi}_2\left(10^{1/2}\pi/a\right)+
 8\tilde{\phi}_3\left(10^{1/2}\pi/a\right)
-2\tilde{\phi}_4\left(10^{1/2}\pi/a\right)$, $A_2= -(23/2)
\tilde{\phi}_1\left(2\pi/a\right)+
 (57/2)
\tilde{\phi}_2\left(2\pi/a\right) -
16\tilde{\phi}_3\left(2\pi/a\right) +
 4 \tilde{\phi}_4\left(2\pi/a\right)
-12\tilde{\phi}_1\left(2^{3/2}\pi/a\right)+9\tilde{\phi}_2\left(2^{3/2}\pi/a\right)-
6\tilde{\phi}_3\left(2^{3/2}\pi/a\right)+3\tilde{\phi}_4\left(2^{3/2}\pi/a\right)$,
and $A_3= - \tilde{\phi}_1\left(4\pi/a\right)+(3/2)
\tilde{\phi}_2\left(4\pi/a\right) -
\tilde{\phi}_3\left(4\pi/a\right)+
(1/4)\tilde{\phi}_4\left(4\pi/a\right)$.

\begin{figure}
\begin{center}

\includegraphics[angle=-90
,width=0.50\textwidth]{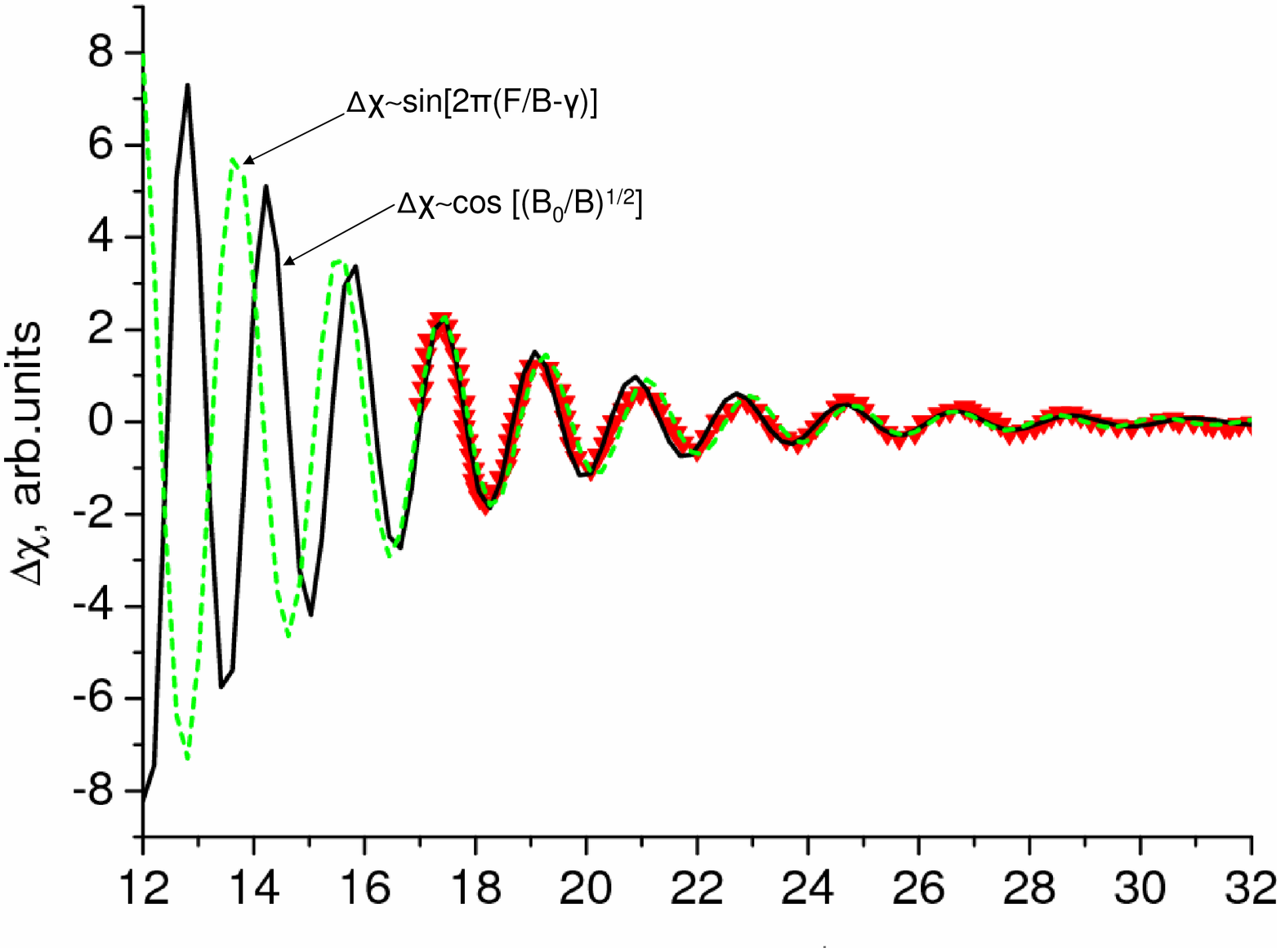}
\includegraphics[angle=-90
,width=0.48\textwidth]{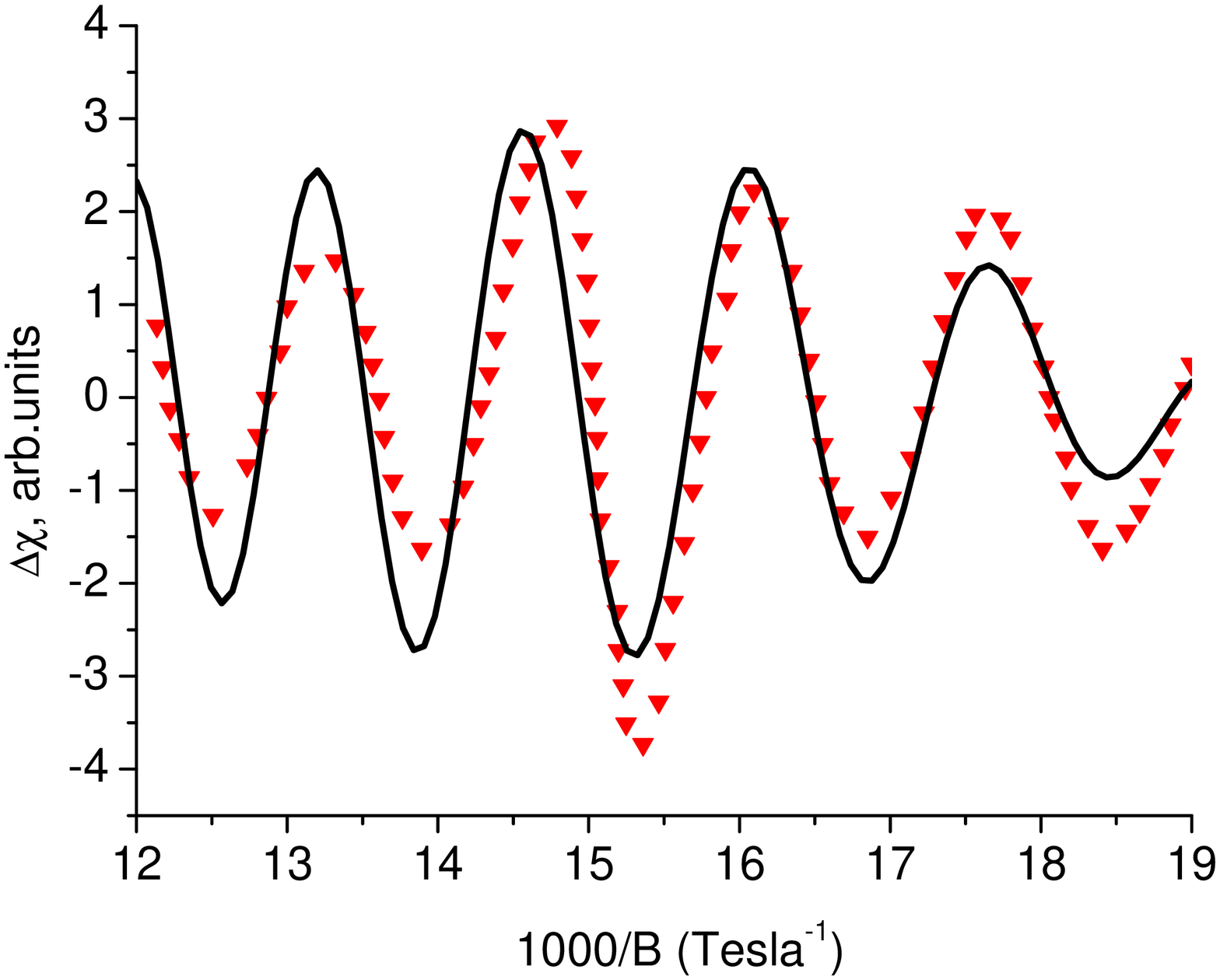}
\includegraphics[angle=-90
,width=0.48\textwidth]{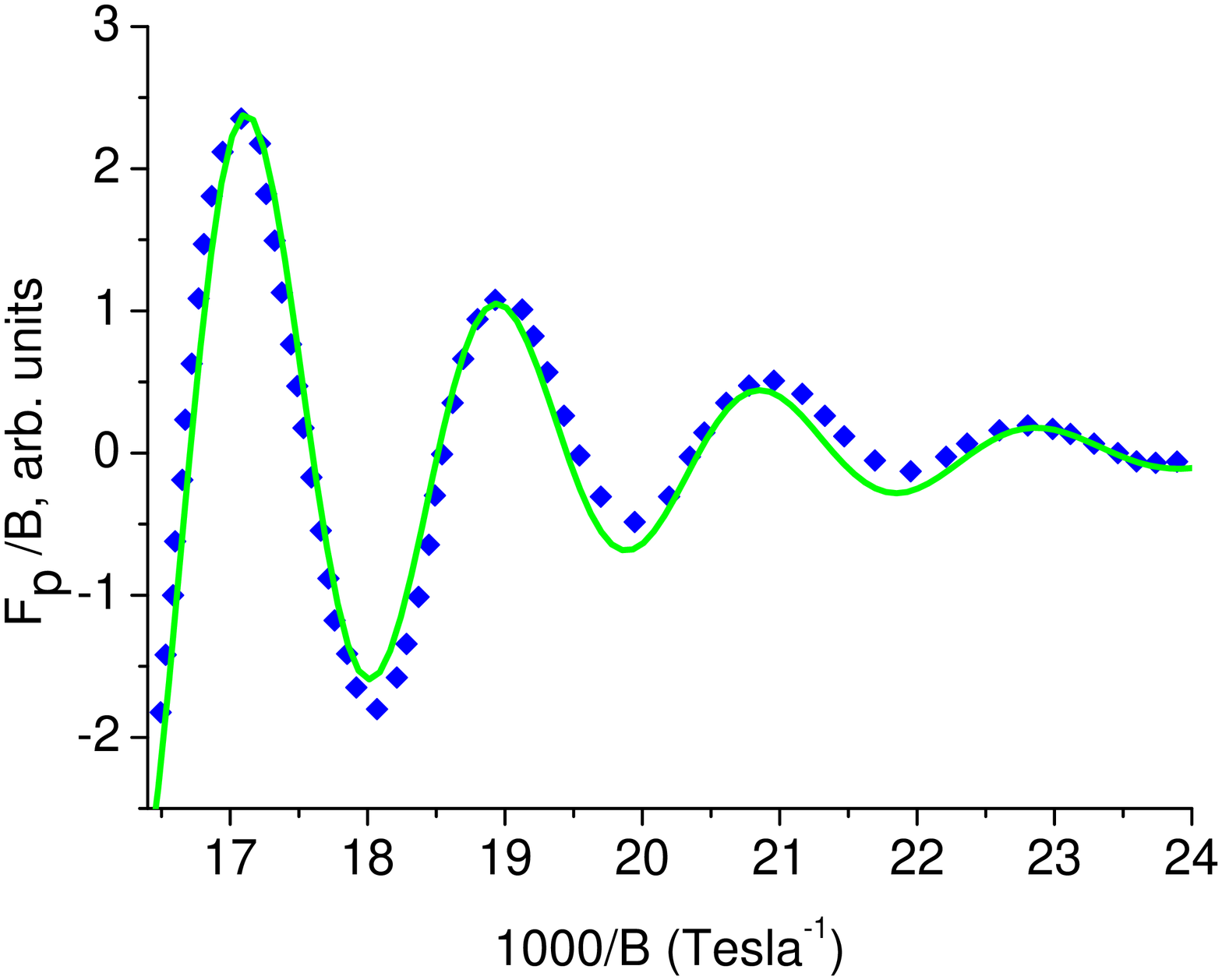}
 \caption{Quantum corrections to
 the vortex-lattice  susceptibility versus $1/B$, Eq.(\ref{osc}) (solid line,
$B_0=1.000\cdot 10^6$ Tesla, $\delta=0.06$) compared with
oscillating susceptibility of YBa$_2$Cu$_3$O$_{6.5}$ (symbols) and
with the conventional normal state oscillations (dashed line)
\cite{proust} at $T=0.4$ K, upper panel. Middle panel: oscillating
susceptibility of YBa$_2$Cu$_4$O$_8$ (symbols \cite{sin}) at
$T=0.53$K compared with the theory (solid line, $B_0=1.190\cdot
10^6$ Tesla, $\delta=0.07$), where  20 percent of the triangular
vortex lattice susceptibility added to the square lattice one. Lower
panel: quantum corrections to the current, proportional to $F_p/B$
($B_0=0.853 \cdot 10^6$ Tesla and $\delta=0.1$, solid line) compared
with the oscillatory part of the Hall resistance in the mixed state
of YBa$_2$Cu$_3$O$_{6.5}$ (symbols \cite{ley}) at 1.5 K.}

\end{center}
\end{figure}

Fluctuations of the pulsed magnetic field and  unavoidable disorder
in cuprates  induce some random distribution of the vortex-lattice
period, $\lambda$. Hence one has to average $\Delta U$ over $R$ with
the Gaussian distribution, $G(R)= \exp
[-(R-\bar{R})^2/\gamma^2]/\gamma \pi^{1/2}$ around an average
$\bar{R}$ with the width $\gamma \ll \bar{R}$. Then  using the
 Poisson summation formula  yields
\begin{eqnarray}
{\Delta U \over{U_0}}= N \sum_{k=0}^{\infty} A_1
e^{-\pi^2k^2\gamma^2/16} \cos (\pi k\bar{R})\cr + A_2
e^{-\pi^2k^2\gamma^2/4} \cos (2\pi k\bar{R})+A_3
e^{-\pi^2k^2\gamma^2} \cos (4\pi k\bar{R}).\label{part2}
\end{eqnarray}

The oscillating correction to the magnetic susceptibility, $\Delta
\chi (B)=-\partial ^2 \tilde{\Omega}/\partial B^2$, is strongly
enhanced due to high oscillating frequencies in Eq.(\ref{part2}).
Since the superfluid has no entropy we can use $\Delta U$ as the
quantum correction to the thermodynamic potential $\tilde{\Omega}$
even at finite temperatures below $T_c(B)$. Differentiating twice
the first harmonic ($k=1$) of the first lesser damped term in
Eq.(\ref{part2}) we obtain
\begin{equation}
\Delta \chi(B) \approx \chi_0  e^{-\delta^2
B_0/16B}\left({B_0\over{B}}\right)^2 \cos (B_0/B)^{1/2},\label{osc}
\end{equation}
where $\chi_0= U_0 SA_1 e^2a^2/4\pi^4\hbar^2$ is a
temperature-dependent amplitude, proportional to the condensate
density squared, $B_0=\pi^3 \hbar/ea^2= 8\pi^3 \hbar/ea_0^2$ is a
characteristic magnetic field, which is approximately $1.1 \cdot
10^6$ Tesla for $a_0 \approx 0.38$ nm, and $\gamma$ is replaced by
$\gamma \equiv \delta \bar{R}$ with the relative distribution width
$\delta$. Assuming that $\xi \gtrsim a $, so that  the amplitude
$A_1$  is roughly $a^2/S$, the quantum correction $\Delta \chi$,
Eq.(\ref{osc}), is of the order of $wx^2/B^2$, where $x$ is the
density of holes per unit cell. It is smaller than the conventional
normal state (de Haas-van Alphen) correction, $\Delta \chi_{dHvA}
\sim \mu/B^2$ \cite{schoen},  for a comparable Fermi-energy scale
$\mu=wx$, since $x \ll 1$ in the underdoped cuprates.

Different from normal state dHvA oscillations, which are periodic
versus $1/B$, the vortex-lattice oscillations, Eq.(\ref{osc}) are
periodic versus $1/B^{1/2}$. They are quasi-periodic versus $1/B$
with a field-dependent  frequency $F=B_0 (B/B_0)^{1/2}/2\pi$, which
is strongly reduced relative to the conventional-metal frequency
($\approx B_0/2\pi$) since $B\ll B_0$, as observed in the
experiments \cite{ley,ban,sin}. The quantum correction to the
susceptibility, Eq.(\ref{osc})  fits well the oscillations in
YBa$_2$Cu$_3$O$_{6.5}$ \cite{proust}, Fig.3 (upper panel).
Importantly, if the vortex lattice has two domains with different
coordination of vortices \cite{ref}, then there are
 two   resonating fields, $B_0$ of the square lattice and $B_1=2B_0/3$ of the triangular lattice, causing  beats in the
 oscillations, as observed in Ref. \cite{sin} at low temperatures in  YBa$_2$Cu$_4$O$_8$(Fig.3, middle panel). A
pinning force on the vortex lattice, $F_p$, due to the checkerboard
modulations  is proportional to $\partial  U/\partial a$. Hence the
oscillating part of the Hall and longitudinal resistivity is
proportional to $F_p/B \propto  \exp{-\delta^2 B_0/16B}(B_0/B)^{1/2}
\sin (B_0/B)^{1/2}$, which fits the oscillatory part of the Hall
resistance \cite{ley} rather well, Fig.3 (lower panel). The
oscillations amplitudes, proportional to $n_s^2\exp(-\delta^2
B_0/16B)$
  decay  with increasing temperature since the randomness of the vortex lattice, $\delta$,
  increases, and
 the Bose-condensate evaporates.

In summary, I propose that the magneto-oscillations in underdoped
cuprate superconductors \cite{ley,ban,sin} result from  the quantum
interference of the vortex lattice and the lattice modulations of
the order parameter, Fig.2, playing a role of the periodic pinning
grid. The magnetic length, $\lambda \gtrsim 5$ nm, remains larger
than the zero-temperature in-plane coherence length, $\xi \lesssim
2$ nm, measured independently, in  any field reached in Ref.
\cite{ley,ban,sin}. Hence the magneto-oscillations are observed in
the vortex (mixed) state well below the upper critical field, rather
than in the normal state, as also confirmed by the \emph{negative}
sign of the Hall resistance \cite{ley}. It is well known, that in
"YBCO" the Hall conductivities of vortexes and quasiparticles have
opposite sign causing the sign change in the Hall effect in the
mixed state \cite{harris}. Also there is a substantial
magnetoresistance \cite{ban}, which is a signature of the flux flow
regime rather than of the normal state. Hence it would be rather
implausible if such oscillations  have a normal-state origin due to
 small electron Fermi surface pockets \cite{proust}  with the
characteristic wave-length of electrons larger than the widely
accepted coherence length. In any case our expression \ref{osc},
describes the oscillations as well as  the standard
Lifshitz-Kosevich formula of  dHvA and SdH effects  \cite{
ley,ban,sin,proust}. The difference of these two dependences could
be resolved in ultrahigh magnetic fields as shown in Fig.3, upper
panel.

 While our theory  utilizes GP-type
equation for hard-core charged bosons \cite{alevor}, the
  quantum interference of vortex and crystal
lattice modulations of the order parameter is quite universal
extending well beyond Eq.(\ref{gp}) independent of a particular
pairing mechanism. It can also take place in the standard BCS
superconductivity at $B < H_{c2}$, but  hardly be observed because
of much lower value of $H_{c2}$ in conventional superconductors
resulting in a very small damping factor, $\propto \exp(-\delta^2
B_0/16B) \lll 1$.

I  appreciate valuable discussions with A. F. Bangura, A.
Carrington, N. E. Hussey, V. V. Kabanov, R. Khasanov,  A.
Paraskevov, I. O. Thomas,  V. N. Zavaritsky, and support of this
work by EPSRC (UK) (grant Nos. EP/D035589, EP/C518365). I am
particularly grateful to Antony Carrington for helping me to correct
an error in the initial fitting of the experimental data.

\end{document}